\documentclass[prb,10pt,twocolumn,superscriptaddress,showpacs]{revtex4}

\usepackage{amsmath}
\usepackage{graphicx}
\usepackage{hyperref}

\newcommand\ket[1]{\left| #1\right\rangle}
\newcommand{\bra}[1]{\left\langle#1\right|}
\def\Phix{\Phi_{\rm x}}
\def\tPhix{{\tilde\Phi}_{\rm x}}
\def\CJ{C_{\rm J}}
\def\omLC{\omega_0}
\begin{document}

\title{Preparation and manipulation of a fault-tolerant superconducting qubit}
\author{Mateusz Cholascinski}
	\affiliation{Institut f\"ur Theoretische Festk\"orperphysik,
	 Universit\"at Karlsruhe, D-76128 Karlsruhe, Germany}
	\affiliation{Nonlinear Optics Division, Institute of Physics,
	  Adam Mickiewicz
	  University, 61614 
	 Poznan, Poland}
\author{Yuriy Makhlin}
	\affiliation{Institut f\"ur Theoretische Festk\"orperphysik,
	 Universit\"at Karlsruhe, D-76128 Karlsruhe, Germany}
	\affiliation{Landau Institute for Theoretical Physics,
	 Kosygin st. 2, 117940 Moscow, Russia}
\author{Gerd Sch\"on}
	\affiliation{Institut f\"ur Theoretische Festk\"orperphysik,
	 Universit\"at Karlsruhe, D-76128 Karlsruhe, Germany}

\date{\today}

\begin{abstract}
We describe a qubit encoded in continuous quantum variables of an
rf superconducting quantum interference device. Since the number of accessible
states in the system is infinite, we may protect its two-dimensional subspace
from small errors introduced by 
the interaction with the environment and during manipulations. We show
how to prepare the fault-tolerant state and manipulate the
system. The discussed operations suffice to 
perform quantum computation on the encoded state, syndrome extraction,
and quantum error correction. We also comment on the physical sources of
errors and possible imperfections while manipulating the system.
\end{abstract}
\maketitle

\section{Introduction}
\label{sec:intro} 

Maintaining quantum coherence is crucial for quantum-information processing.
Coherence of any quantum system is gradually suppressed due to unwanted
interactions with the environment.  Among proposed realization of qubits, 
solid-state devices appear particularly promising due to the scalability and 
ease of integration in electronic circuits, but their operation
requires keeping them coherent, a potentially strong problem due to
the host of microscopic   
modes.  One of the achievements of
quantum-information theory is the development of quantum-error-correction
techniques, which allow one to suppress the effect of the environment on the
software level, i.e. by running the appropriate quantum code to eliminate 
errors.\cite{NielChu} The standard approach protects against large errors that
occur rarely.\cite{gottesman, calderbank, rains}  Alternative methods were proposed by
\textcite{GotPreKit}, where the {\em shift-resistant} codes
protect against errors that occur continuously but are weak.
Specifically, they analyzed the situation when a qubit is embedded in an
infinite-dimensional Hilbert space of a physical device with
a continuous degree of freedom. Continuous-variable quantum codes
have been further developed for both qubits and qudits ($d$-dimensional analog of
qubits).\cite{braunstein0,barlett,pirandola,barlett1,sanders} The
continuous-variable codes became also a framework for discussion of
quantum key distribution \cite{got} or quantum teleportation of
continuous quantum variables.\cite{braunstein2}.

In the original work,\cite{GotPreKit} the authors focused on encoding a qubit in an
oscillator and developed codes which protect against small shifts in the
canonical variables of the oscillator.  They also discussed implementation of
this approach in optical systems.  A universal set of quantum logic gates and
error-correction steps may be realized using linear optical operations,
squeezing, homodyne detection, photon counting, and nonlinear couplings.  While
some of these steps are easily realized with optical means, the encoding and
certain gates from the universal set, which require photon counting and
nonlinear couplings, are difficult to implement (alternative method of
encoded state preparation has been discussed in
Ref.~\onlinecite{travaglione}).

Here, we suggest the implementation of the shift-resistant codes in superconducting
devices.  We show that their physical properties simplify the implementation of
the difficult steps mentioned above. 
In these systems one may 
use the charge \cite{makhlin0,Pashkin1,Pashkin2} or the conjugate
phase degree of freedom \cite{martinis, cooper} to store and process
quantum information.  Typically, one adjusts the parameters such that
the ground 
state is (almost) doubly degenerate, and at low energies the system
reduces to a 
qubit.  Here we consider a different approach, in which a continuous degree of 
freedom is used:  the magnetic flux through the loop of an rf-SQUID
(superconducting quantum interference device) (a
superconducting loop interrupted by a Josephson junction) and the
conjugate charge. The Josephson coupling can be tuned if one replaces
the junction with a dc-SQUID.\cite{Tinkham} For a turned-off
Josephson coupling,  
the system behaves as a harmonic oscillator, and one can encode a
qubit in its Hilbert space and manipulate the qubit's state. The
specific useful  
features of this design are related to the periodic flux dependence of certain 
physical properties: this allows one to control the Josephson coupling and to 
monitor the magnetic flux modulo the flux quantum, thereby projecting out a 
comblike state needed for encoding of the qubit.

The goal of our work is twofold. On one hand, we suggest an implementation of 
certain steps needed for shift-resistant codes, which are hard to
implement with 
optical means. On the other hand, one may think of implications of our results 
for the long-term strategy of quantum computing in superconducting systems. 
While 
fabrication of larger circuits should be relatively straightforward,
it will be  
necessary to protect them against decoherence. Although the approach
we discuss  
here imposes certain constraints on the system parameters and requires 
complicated operation procedures, the alternative of standard 
quantum-error-correction codes requires building circuits with many auxiliary 
qubits and performing complicated series of logic gates during error 
detection and recovery (cf. Ref.~\onlinecite{top-qubits}). Here, we demonstrate that the 
ideas of continuous-variable codes may be, in principle, implemented in 
superconducting circuits, underline advantages of these devices but do not 
optimize the design, and only superficially consider  constraints on
the circuit parameters.

We begin with a short description of the shift-resistant codes in 
Sec.~\ref{sec:shift}. In
Sec.~\ref{sec:logical}, we briefly describe the physical system
representing the qubit and discuss the proposed implementation of the 
difficult steps, i.e., of the encoding procedure and the quantum
gates, including
those needed for error correction. In Sec.~\ref{sec:manipu}, we discuss, for 
completeness, further necessary steps such as squeezing and
translations in the  
phase space and two-qubit operations. We then comment on the 
error models relevant for practical devices and discuss constraints on the 
circuit parameters.

\section{Shift-resistant codes for $LC$ oscillator}
\label{sec:shift}

While qubit is the simplest nontrivial quantum system, many physical
systems offer the opportunity to utilize many 
levels and often a continuous spectrum to process quantum
information. \textcite{GotPreKit} suggested 
to encode  
a logical qubit in a system with a continuous degree of freedom, an
oscillator.   
They described error-correcting codes, which protect the state of the qubit 
against perturbations that cause weak diffusion of the position and
momentum of  
the oscillator. Here, we briefly describe this proposal in the language of an 
$LC$ circuit.

Consider an $LC$ oscillator, with the Hamiltonian
\begin{equation}
  H = \frac{\Phi^2}{2 L} + \frac{Q^2}{2 C} \,.
  \label{eq:Mwke}
\end{equation}
Here, the dynamical variables are the magnetic flux $\Phi$ and the
conjugate
charge $Q$.  For convenience, we use below the dimensionless flux and charge 
variables,
\begin{equation}
  \Phi'  = {1 \over \sqrt{\pi}} \left({C\over L}\right)^{1/4} \Phi \,,
  \quad  Q' =   {1 \over \sqrt{\pi}}\left({L\over C}\right)^{1/4} Q \,,
  \label{eq:Mqhe}
\end{equation}
and omit the primes.
The rescaled variables satisfy the standard commutation
relation
\begin{equation*}
  [\Phi , Q] = {i \over \pi }.
\end{equation*}

In the proposal of \textcite{GotPreKit} the codewords (the states that encode 
the basis logic states of the qubit) are comblike superpositions, both in the 
flux and charge representations,
\begin{eqnarray}
  \nonumber
  |\bar{0} \rangle &=& \sum\limits_{s = - \infty }^{\infty } |\Phi =
  2s \rangle = \sum\limits_{m = -\infty }^{\infty } |Q = 
  m \rangle \,, \\
  |\bar{1} \rangle &=& \sum\limits_{s = - \infty }^{\infty } |\Phi =
  2s + 1 \rangle = \sum\limits_{m = -\infty }^{\infty
    }(-1)^{m} |Q = m \rangle \,.
  \label{eq:codewords}
\end{eqnarray}

For the purpose of error correction one measures the value of
$\Phi$ mod $1$.  Such a measurement provides information on the possible
error, a shift of the comblike states [Eq.~(\ref{eq:codewords})], but does not
distinguish between the combs peaked at even and odd flux values,
$\ket{\bar 0}$ 
and $\ket{\bar 1}$.  The observed shift is compensated by the inverse flux
shift, which moves the comb structure to the closest integer.  This procedure
can correct sufficiently small flux-shift errors, $\Delta\Phi<1/2$.  Similarly,
one can correct small shifts, $\Delta Q<1/2$, in the charge variable
$Q$. If the  
protection from rare large errors is also desired, these codes can be 
concatenated with the standard error-correction codes.\cite{GotPreKit}

The states [Eq.~(\ref{eq:codewords})] define a two-dimensional subspace
protected from  
decoherence. In the following sections, we will show how one can
prepare such states, store them (in the 
oscillatory regime with the Josephson coupling turned off),
implement quantum logic gates and error-correction steps
using the control over the Josephson coupling, the flux bias, and inductive 
coupling between qubits. 

\section{Encoding the qubit and logic gates}
\label{sec:logical} 

\subsection{Oscillator and the identity operation}

\begin{figure}[h] 
\centerline{\resizebox{0.6\columnwidth}{!}{\rotatebox{0}
{\includegraphics{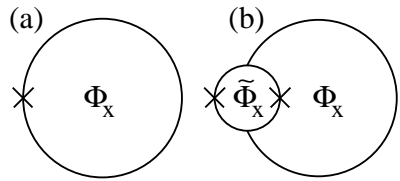}}}}
\caption{\label{fi:fluxqubit} The simplest flux qubits (Ref.~\onlinecite{makhlin})
  can be used to encode a qubit in their continuous variables. (a) The
  rf-SQUID, a simple loop with a Josephson junction, and (b) rf-SQUID with
  tunable Josephson coupling.} 
\end{figure}
The Hamiltonian of an rf-SQUID  with a tunable Josephson coupling [shown in 
Fig.~\ref{fi:fluxqubit}(b)] reads
\begin{eqnarray}
\nonumber
H &=& \pi \omega _0 \left[ \frac{(\Phi - \Phix)^2}{2} + \frac{Q^2}{2}
  \right] \\
&&  - E_J(\tPhix) \cos \left[2 \pi^{3/2} (L/C_J)^{1/4}
  \frac{\Phi}{\Phi_0}\right]\,.
\label{eq:hamiltonian}
\end{eqnarray}
Here, the first two terms, which form the Hamiltonian of an oscillator, are the 
magnetic energy, controlled via the external flux bias $\Phix$ and the 
inductance $L$ of the loop, and the charging energy (depending on the
parameters  
the capacitance $\CJ$ may reduce to that of the junction or involve
the geometry  
of the whole loop). $\omLC= (LC_J)^{-1/2}$ is the frequency of the
$LC$ oscillator. The last term is the Josephson energy, with magnitude 
\[
E_J(\tPhix) = 2 E_J^0 \cos \left[\pi^{3/2} (L/\CJ)^{1/4} \tPhix
  / \Phi_0 \right] \,,
\]
controlled by the flux $\tPhix$ through the small loop of the dc-SQUID.

We observe the qubit's states [Eq.~(\ref{eq:codewords})] in the interaction 
picture of dynamics. In this approach, the state of the qubit is conserved if no 
manipulations are performed (although in the Schr\"odinger representation, the 
Hamiltonian generates evolution), thus providing the identity
operation. Specifically, consider the system in the oscillatory
regime, when the  
Josephson coupling and the externally applied flux $\Phix$ are tuned to zero.
Then for an arbitrary state, the Hamiltonian [Eq.~(\ref{eq:Mwke})], with
the parabolic potential centered at the origin, generates rotation (of the 
density matrix in the Wigner form) of the state with the frequency
$\omLC$ in the phase space
$\Phi$-$Q$ around the origin.
In other words, the amplitudes $\Phi_R$, $Q_R$ in the rotating frame 
(interaction representation), which are related to $\Phi$ and $Q$ as
\begin{equation}
  \left(
    \begin{array}{c}\Phi_R\\Q_R\end{array}
  \right) = 
  \left(
    \begin{array}{lr}
      \cos \omLC t & - \sin \omLC t   \\
      \sin \omLC t & \cos \omLC t
    \end{array}
  \right) 
  \left(
    \begin{array}{c}\Phi\\Q\end{array}
  \right)\,,
  \label{eq:M8ie}
\end{equation}
are integrals of motion, $\dot{\Phi }_R = \dot{Q}_R = 0$. Below we discuss 
quantum logic transformations in this reference frame.  Note that performing
operations in the rotating frame implies in practice that one has to
keep track of the phase  
of the oscillator at frequency $\omega_0$ and perform all steps relative to 
this reference signal.

\subsection{Encoding}

Using the possibility to measure the flux mod $\Phi_0$, one can prepare a 
code state [Eq.~(\ref{eq:codewords})], $\ket{\bar 0}$or $\ket{\bar 1}$, by first 
preparing a state with a wide uniform distribution of flux and then
measuring the  
flux value. As a result, one obtains a comblike state, with equidistant peaks 
and an offset of the structure, given by the result of the flux
measurement. One  
performs a flux shift to compensate this offset and arrives at a codeword 
[Eq.~(\ref{eq:codewords})]. Possible methods to implement these steps are discussed 
below.

Preparation of a wide flux distribution may be realized by a number of methods. For 
instance, one may let the oscillator relax to the ground state in a
narrow well,  
with the inductance decreased by (for instance, slowly) switching on a strong 
Josephson coupling. This creates a ``squeezed'' state, with a narrow flux 
distribution and a wide charge distribution. When the Josephson coupling is 
turned off, this distribution in the $\Phi$-$Q$-plane starts rotating
about the  
origin and after a quarter of the oscillator period, $\pi/(2\omLC)$, a wide 
distribution of flux is reached, $\sim\int d\Phi
\ket{\Phi}$. Alternatively, one  
can prepare such a state by squeezing the ground state of the oscillator (see 
below).

\begin{figure}[t] 
    \centerline{\resizebox{0.55\columnwidth}{!}{\rotatebox{0}
        {\includegraphics{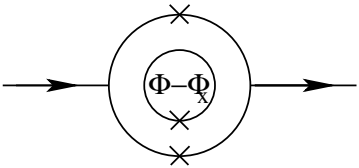}}}}
    \caption{\label{fi:preparation} A possible method of the
       inductive meter-qubit coupling. In this geometry (when the
       qubit is placed inside the meter), the coupling is strong, and
       $\lambda \approx 1$ . For simplicity, the qubit is
      shown without the smaller, dc-SQUID loop.}
\end{figure}

Now, to single out a comblike structure out of this wide distribution,
one is to  
perform a measurement of the flux value up to a multiple of a certain offset. 
This can be achieved by coupling the rf-SQUID inductively to a
readout dc-SQUID  
and reading out the critical current of the latter. This critical current is a 
periodic function of the total magnetic flux in the measuring SQUID's loop,
\[
    I_{\rm c} = 2I_{{\rm c}0}
    \cos\left[ \pi \frac{\Phi_{\rm mx} + \lambda (\Phi-\Phix)}{\Phi _0} 
\right]\,, 
\]
where $\Phi_{\rm mx}$ is the external flux bias of the measuring SQUID and
the coupling $\lambda$ depends on the self-  and mutual inductances of
the loops, i.e., on the geometry of the SQUID's. For the particular geometry
depicted in Fig.~\ref{fi:preparation}, the parameter $\lambda $ can be
close to $1$ (our scheme is valid for different scenarios and values
of $\lambda$ as well). 

Reading out the value of the critical current, one projects out a comblike 
state,
\begin{equation}  
    P_\alpha = \sum\limits_{s = - \infty }^{\infty}
    \ket{\Phi = s \Phi_{\rm p} + \alpha}
    \bra{\Phi = s \Phi_{\rm p} + \alpha},
\label{eq:Pal}
\end{equation}  
with a period $\Phi_{\rm p}= \Phi_0/\lambda$ (if only the absolute value of 
critical current is measured and the sign is not resolved). $\alpha $
here is the initial displacement of the state from zero. (In fact,
one would  
obtain a superposition of two comblike states with different offset values
$\alpha $ since
two series  
of delta peaks correspond to each value of the critical current; one method to 
leave only one of these is to read out the critical current again
after a small  
shift of the flux bias $\Phi_{\rm mx}$.)

Starting from the state projected by Eq.~(\ref{eq:Pal}) from the wide flux
distribution, one could compensate for the shift 
$\alpha$ by performing the inverse flux shift and also change the
period of the comb structure by squeezing (see the next section) to arrive
at one of the basis states, $\ket{\tilde 0}$ or $\ket{\tilde 1}$. As
we shall see later, for
the purpose of performing one-qubit gates it
is convenient to tune the peak separation to the superconducting flux
quantum, $\Phi _0$ (possibly by squeezing). However, according to the
definition of the encoded states [Eq.~(\ref{eq:codewords})] and in the units
defined in Eq.~(\ref{eq:Mqhe}) the peak spacing in the state $|\bar{0}
\rangle $ equals $2 (\pi ^2 L/\CJ)^{1/4} $. Comparison to the flux
quantum gives an extra
constraint on the ratio $L/C_J$,\footnote{The choice of units following
from Eq.(\ref{eq:Mqhe}) makes the dimensionless variables fully
symmetric under exchange in the Hamiltonian Eq.(\ref{eq:Mwke}). It is thus the
only one for which the oscillatory evolution is rotation around a
circle (not ellipse) in the phase space. Conservation of the distance
is crucial for the one-qubit operations and hence the constraint.} 
\begin{equation}
  \left({L \over C_J}\right)^{1/4} =  \Phi _0 / (2 \sqrt{\pi }).
  \label{eq:condit}
\end{equation}
With this condition satisfied, the Josephson term in the Hamiltonian of the system
[Eq.~(\ref{eq:hamiltonian})] becomes periodic in $\Phi $ with the period $2$.

\subsection{Quantum gates}

If the Josephson coupling of the SQUID loop is nonzero, the
Hamiltonian in the interaction picture has for $\Phi _x = 0$ the
following form
\begin{equation}
  H = - E_J (\tPhix) \cos \left[\pi \Phi_R \cos \omega_{0} t +
    \pi Q_R \sin \omega_{0} t \right].
  \label{eq:M9ie}
\end{equation}  
For the times $t_z = k \pi / \omega_{0} $ and $t_x = \pi (k + 1/2) /
\omega_{0} $ ($k$ being an integer), it reduces in the code subspace
to
\begin{subequations}
\begin{eqnarray}
  \label{eq:siz}
    H(t_z) &=& - E_J (\tPhix) \sigma _z, \\
    H(t_x) &=& - E_J (\tPhix) \sigma _x.
\label{eq:six}
  \end{eqnarray}
\end{subequations}
Here, the Pauli matrices act on the encoded states $|\bar{0} \rangle $
and $|\bar{1} \rangle $ like on the spin states ``up'' and ``down''
along the $z$ direction respectively.
Using short pulses (the duration $\tau $ should satisfy $\tau \omega
_{0} \ll 1$) of the magnetic flux $\tPhix$ at the times
$t_{x(z)}$, we can perform small phase shifts between the encoded
states or induce transitions between them. Sufficiently large
shifts, accumulated during many pulses, give rise to simplest
one-qubit operations that generate the group of all possible unitary
transformations in the encoded subspace.\cite{barenco}

Certainly no operation can be performed instantaneously. So, apart from
the instantaneous Hamiltonian corresponding exactly to the times
$t=t_{x(z)}$, we should analyze the time dependence of the
Hamiltonian near $t_{x(z)}$ to
evaluate the error introduced by the finite rotation during the
operation. Let us consider the Hamiltonian [Eq.~(\ref{eq:M9ie})] at
times close to $t_z$. Expansion to the quadratic terms in $\omega_{0} \tau$
gives
\begin{eqnarray*}
  \nonumber
  H &\approx& - E_J (\tPhix) \left\{ \cos [\pi \Phi _R] - \pi
  Q_R \sin [\pi \Phi _R] \omega_{0} \tau - \right.\\
  && \left. - [(\pi Q_R)^2\cos (\pi \Phi _R) - (\pi \Phi _R) \sin (\pi
  \Phi _R)] 
  (\omega_{0} \tau )^2 \right\}.
\nonumber 
\end{eqnarray*}
The error is generated by the time-dependent part. Since for the code space
\[
  \sin (\pi \Phi _R) |\bar{0} \rangle = \sin (\pi \Phi _R) |\bar{1}
  \rangle = 0,
\]
the error is generated by at least second order terms in $\omega _0
\tau $, and
for $\omega_{0} \tau \ll 1$ it is small. In addition, as shown in
Sec.~\ref{imper}, errors generated by quadratic terms in $Q$ can be
corrected using the error-correcting routines. (The same property
holds at times close to $t_x$.)

To complete the set of universal quantum gates, we need also a
many-dimensional (in this case continuous) equivalent of CNOT, the SUM
gate, which transforms the variables of two coupled qubits according to 
\begin{eqnarray*}
  \nonumber
  \mbox{SUM}: && \Phi _1 \rightarrow \Phi _1, \quad Q_1 \rightarrow
  Q_1 - Q_2, \\
  && \Phi _2 \rightarrow \Phi _1 + \Phi _2, \quad Q_2 \rightarrow Q_2.
 \nonumber 
\end{eqnarray*}
Here, the indices number the coupled qubits. 
This gate belongs to the so-called {\em symplectic group} and may be
realized using (in quantum-optical setting) phase shifters, squeezing, and
beam splitters \cite{GotPreKit,reck} - elements that are
accessible also for $LC$ oscillators, as we will show in the next
section. The SUM gate reduces in the code subspace to CNOT. Its
continuous nature is, however, crucial during the syndrome
extraction.

\section{Manipulating the system}
\label{sec:manipu}

So far, we presented the computational steps that are difficult
with quantum-optical elements and much more natural with the
Josephson-junction 
systems. To make the discussion complete, let us now discuss the
operations that are necessary for the symplectic operations and can be
performed on arbitrary wave function: 
squeezing, phase shifts, translations in the $\Phi $-$Q$ plane, and
inductive coupling of two oscillators. In this section, all operations are
described in the oscillatory regime $E_J(\tPhix) = 0$.

\subsection{Squeezing and phase shifts}

The scaling factors in Eq.~(\ref{eq:Mqhe}) define dimensionless
variables for which the Hamiltonian is parametrized by only one real
parameter, the energy scale $\omega _0$. Since the scaling depends on
the ratio $L/\CJ$ and the frequency $\omega _0$ on the product of $L$
and $\CJ$, modification of either of the parameters ($L$ or $\CJ$)
influences the system behavior in two ways - the oscillation period
changes and the evolution (rotation) path in the phase space is squeezed in one
direction. In particular, suppose that there is an
extra capacitance in parallel to the junction that can be switched
instantaneously on and off, so that the total capacitance may equal
$C_J$ or $\lambda C_J$. Switching at some instant of time [for
simplicity we assume that this moment corresponds to $t=0$ in
Eq.~(\ref{eq:M8ie})] from $\CJ$ to $\lambda \CJ$ rescales the
variables, $\Phi = \Phi _R 
\rightarrow \lambda ^{1/4} \Phi = \lambda ^{1/4} \Phi_R, Q = Q_R
\rightarrow \lambda ^{-1/4} Q = \lambda ^{-1/4} Q_R$, and
modifies the frequency, $\omega _{0} \rightarrow \omega _{1} = \lambda
^{-1/2} \omega _{0}$. Using this property, we can perform squeezing of
arbitrary wave function. However, according to Eq.~(\ref{eq:condit}), there is a
constraint relating the system parameters to the value of the flux
quantum and we 
have to switch the capacitance back to the original value before
performing further steps, like the single-qubit operations. The second
switching (back from $\lambda \CJ$ to $\CJ$) should be performed after
a time delay, corresponding to an exchange of the role of the
variables. To be more specific suppose that we switch the capacitance
from $\CJ$ to $\lambda \CJ$ at $t=0$, switch it back at $t= \pi /2
\omega _1$ (quarter of the full oscillation), and observe the effect
after the period of oscillation is 
completed (then the laboratory and rotating frames coincide and the
effect is identical, which simplifies the analysis). The evolution in
the Heisenberg picture of dynamics (in the laboratory frame) is described by
the inverse of Eq.~(\ref{eq:M8ie}),
\[
  \left(
    \begin{array}{c}\Phi\\Q\end{array}
  \right) = 
  U^{-1}(\omega _{0(1)}, t)
  \left(
    \begin{array}{c}\Phi_R\\Q_R\end{array}
  \right)\,,  
\]
where $U^{-1}$ is the inverse of the rotation matrix in Eq.~(\ref{eq:M8ie}).
In this picture (and frame), the scaling of variables corresponding to
the switching from $\CJ$ to $\lambda \CJ$ is described by a
diagonal matrix $S$ with the elements $\lambda ^{1/4}$ and $\lambda
^{-1/4}$. Thus, the entire procedure is described by the following operation:
\begin{eqnarray*}
  U_{squeeze} &=& U^{-1}(\omega _0, 3 \pi / 2 \omega _0) S^{-1} U^{-1}(\omega _1, \pi
  /2 \omega _1) S \\ &=& \left(
  \begin{array}{lr}
    \lambda ^{1/2} & 0   \\
    0 & \lambda ^{-1/2}
  \end{array}
  \right).
  \label{eq:yva}
\end{eqnarray*}
The interaction-picture result is 
\[
  U_{squeeze}\left(
  \begin{array}{c}
    \Phi _R \\
    Q_R
  \end{array}
  \right) = \left(
  \begin{array}{c}
    \lambda ^{1/2} \Phi _R\\
    \lambda ^{-1/2} Q_R
  \end{array}
  \right).
\]
This equation clearly describes squeezing of an arbitrary state. Note
that after we change the capacitance of the qubit, the frequency
is modified as well. This can be used in implementing another element -- the
shifter of the relative phase 
between two oscillators. If we do not want the state to be modified but
only the evolution advanced (or delayed), we should switch the
capacitance to $\lambda C_J$ for the full period $2 \pi / \omega
_{1}$.

\subsection{Translations in the phase space}

The evolution in the interaction picture (the rotating frame) is
trivial, provided that the external flux $\Phix $ is zero or at least
constant. Once we switch it 
to a finite value, the physical center of rotation in the phase space
shifts by $\Phix $, and to simplify the
description 
again we would have to shift the frame as well. However, square
pulses of the flux $\Phix$ give us the possibility to perform another useful
and necessary in error-correcting routine transformation -- translations in
the phase space (in the $\Phi $-$Q$ plane).

To avoid these noninertial effects of the rotating frame, we describe
this procedure in the laboratory frame (it lasts, however, for the full
oscillation period, and like in the case of squeezing the effects are
in both frames identical after the operation is over). Suppose that we
begin with a state described by the wave 
function $\psi _0 (\Phi )$ in the potential centered at the
origin ($\Phi _x = 0$). At the time $t=0$, we turn on the external flux
$\Phi_x = - \beta/2$. After the time $t = \pi / \omega_{0} $,
which corresponds to a half of the oscillation, the state is
transformed into
$\psi _1(\Phi ) =
\psi _0 (- \Phi +\beta)$. Then, we turn the external flux off
and after the oscillation is completed, we obtain $\psi _2( \Phi ) = \psi
_0 (\Phi - \beta)$ -- the initial state shifted by
$\beta$ in $\Phi $. To shift the state in the $Q$ direction we need to
perform the same procedure on the Fourier-transformed state,
  i.e., to delay the initial moment of operation by $\pi / 2
\omega_{0}$.

\subsection{Inductive coupling of oscillators}

\begin{figure}[h] 
    \centerline{\resizebox{0.4\textwidth}{!}{\rotatebox{0}
        {\includegraphics{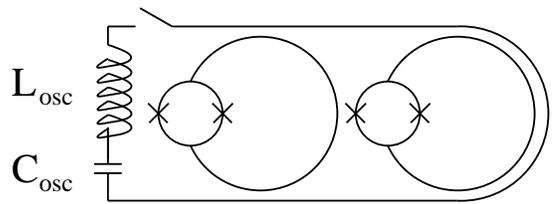}}}}
    \caption{\label{fi:coupling} Inductive coupling of two qubits is
        realized with the Josephson energy tuned to zero. It is thus
        the coupling of two $LC$ oscillators. In the rotating frame of
        reference, this coupling provides equivalent of quantum-optical
        beam splitter (see the text for explanation).}
  \end{figure}

  Two qubits may be coupled inductively using an
  additional $LC$ circuit (see Fig. \ref{fi:coupling}).
  The Hamiltonian of the $LC$ oscillator and the qubits (for a moment we
  turn back to the natural units) is
  \begin{eqnarray}
    \nonumber
    H &=& H_1 + H_2 + \Phi ^2 / 2 L_{osc} + Q^2 / 2 C_{osc} - VQ, \\
    V &=& \sum\limits_{i}^{} M_i \dot{\Phi }_i / L,
    \label{eq:Mspa}
  \end{eqnarray}  
  where $H_1$ and $H_2$ are oscillatorlike Hamiltonians of the form
  of Eq.~(\ref{eq:Mwke}), $Q$ is the charge on the leads of the capacitor
  $C_{osc}$, and $\Phi $ is the flux through the external $LC$ circuit. If the 
  frequency of the oscillator is much bigger than the frequency of the qubits
  $\omega _{LC} = 1 / \sqrt{L_{osc}C_{osc}} \gg \omega_{0} =
  1/\sqrt{L \CJ}$ (in the limit $C_{osc}\rightarrow 0$), we find that 
  the coupling circuit remains in its ground state and the
  interaction (mediated by virtual excitations of the coupling circuit)
  is described by the term $-C_{osc} V^{2} / 2$. 
  Since $\dot{\Phi }_i = Q_i / C_J$, the interaction part of the
  Hamiltonian for two qubits has the form
  \begin{equation}
    H_{int} = - \sum\limits_{i \neq j} {C_{osc} M_i M_j \over L^2
      C_J^2} Q_i Q_j \equiv - K_{12} Q_1 Q_2,
    \label{eq:coupling}
  \end{equation}
  so that after integrating out the
  oscillator's variables, the Hamiltonian of the interacting system equals
\[  
    H = H_1 + H_2 - K_{12} Q_1 Q_2.
\]
  (Here, the parameters in $H_{1(2)}$ are slightly different than in
  Eq.~(\ref{eq:Mspa}) - from the term $-C_{osc} V^{2} / 2$ we obtain
  the interaction part of the Hamiltonian (\ref{eq:coupling}) as well
  as terms quadratic 
  in $Q_{1(2)}$, which formally slightly modifies the frequency of the
  qubits. In any case, this does not influence the general shape of
  our results - the phase shifts between groups of qubits can be
  compensated using the previously discussed techniques.) We can rewrite
  this Hamiltonian in the rotating frame of 
  reference,  
  \begin{eqnarray*}
    \nonumber
    H &&= - {K_{12} \over 2} \left[Q_{R 1}Q_{R 2} (1 + \cos 2
    \omega_{0} 
    t) + \right. \\ &&
    \left. \Phi_{R 1} \Phi_{R 2} (1 - \cos 2 \omega_{0} t) +
    (Q_{R 1} 
    \Phi _2 + Q_{R 2} \Phi _1) 
    \sin 2 \omega_{0} t\right].  
\nonumber 
  \end{eqnarray*}
    Since over time scales much longer than $1/\omega_{0}$ the effect of the
    oscillating terms averages out, we finally arrive at
\[     
      H = - {K_{12} \over 2} \left(Q_{R 1}Q_{R 2} + \Phi _{R 1} \Phi
      _{R 2} \right).
\]  
    Now, we solve the Heisenberg equations of motion for the operators with
    the initial conditions $\Phi _{1(2)}(0) = \Phi _{1(2)}^{0}$, $
    Q_{1(2)}(0) =Q_{1(2)}^0$ (for simplicity we omit here the $R$
    subscripts), and we arrive at
    \begin{eqnarray*}
      \nonumber
      \Phi _{1(2)} (t) &=& \Phi _{1(2)}^{0} \cos {K_{12} t\over 2} +
      Q_{2(1)}^{0} \sin {K_{12} t\over 2},  \\
      Q_{1(2)} (t) &=& - \Phi _{1(2)}^{0}\sin {K_{12} t\over 2} +
      Q_{2(1)}^{0} \cos {K_{12} t\over 2}.
\nonumber 
    \end{eqnarray*}  
    If we make further the substitution (which is done in practice by
    advancing the second oscillator by $\pi / 2$),
    \[    
    \Phi _2 \rightarrow - Q_2, \quad Q_2 \rightarrow \Phi_2,
    \]
    we obtain equations describing the action of a beam splitter
    \footnote{In 
    quantum optics the oscillator is simply a laser beam. The oscillatory
    variables $q$ and $p$ correspond to different quadratures of the
    electromagnetic field}
    with time-dependent reflectivity and transmittance,
    \[    
    \sqrt{R} = \cos {K_{12} t\over 2}, \quad \sqrt{T} = \sin {K_{12}
      t\over 2}.
    \]  
    
    This element completes the set of all necessary operations.

\section{Practical considerations}\label{imper}

\subsection{Error models and error recovery}
Errors that can be corrected using shift-resistant quantum codes are
translations that leave the state out of the code subspace but still
not too far from the initial state. The errors that can occur are due
to the following interaction with the environment:
\begin{equation}
  |\overline{x} \rangle _{qb} |0 \rangle _E \rightarrow \left(e^{i Q
      \alpha }e^{i \Phi \beta } |\overline{x} \rangle \right)
  |\alpha , \beta  \rangle _E.
  \label{eq:Mske}
\end{equation}  
Certainly, almost no physical interaction is of this kind, but it can
be expanded in terms of such translations,
\begin{equation}
  |\overline{x} \rangle _{qb} |0 \rangle _E \rightarrow 
  \int d \alpha' d \beta' C(\alpha' , \beta' , t) \left(e^{i Q
      \alpha' }e^{i \Phi \beta' } |\overline{x} \rangle \right)
  |\alpha' , \beta'  \rangle _E
  \label{eq:Mtke}
\end{equation}  
The syndrome extraction procedure gives then one pair of real
parameters $(\alpha , \beta )$ only, projecting the state onto
Eq.~(\ref{eq:Mske}). Effectiveness of the error recovery depends in this
case on the amplitude $C$. We may treat the function $|C|^2$ as the
probability distribution of possible errors (or equivalently $C$ as
two-dimensional wave function). If the uncertainties $\Delta \alpha' $
and $\Delta \beta' $ are small after the interval between two
error-correcting routines, the states will be
recovered with high fidelity.

The interaction with the environment can be also written in the
operator-sum representation (more convenient for our purposes)
\begin{eqnarray}
  \rho (t) &=& \sum\limits_{i} M_i \rho (0) M ^{\dagger}_i,
  \label{eq:superop} \\ 
  M_i &=& \int d \alpha d \beta C_i (\alpha , \beta , t) e^{i \beta
  \Phi}  e^{ - i \alpha Q}.
  \label{eq:Mvie}
\end{eqnarray}
In this case, if the error superoperator [acting on the qubit density
matrix as in Eq.~(\ref{eq:superop})] has support in operators that can be
expanded in terms of small shifts the error correction will be
effective.

This is indeed the case for some classes of physical errors. Amplitude
damping of harmonic oscillator and a class of unitary errors like
over-rotation or under-rotation have been described in 
Ref.~\onlinecite{GotPreKit}. Such errors can occur easily from delay or
advance in the pulse application, for instance during the
single-qubit manipulations, where we need to keep track of the
time evolution. However, the physical sources of errors 
can be different in quantum optical and superconducting
systems. While in optics imperfections 
in operations lead mainly to amplitude damping (for instance
absorption of light on the beam splitters and phase shifters),
superconducting elements 
are dephased by the Josephson junctions. So, to make
our discussion more explicit, we consider (apart from the models
presented already in
Ref.~\onlinecite{GotPreKit}) also the phase damping of 
the $LC$ oscillator. 

Specifically, let us consider the full Hamiltonian of the rf-SQUID, where 
the external flux ($\Phi _x$ and $\tPhix$) may be
slightly fluctuating 
\begin{eqnarray}
  \nonumber
  H &=& - 2 E_J^0 \cos \left[{\pi \tPhix + \pi \delta \tPhix (t)\over
    2}\right]  
\cos
  \left( \pi \Phi \right) \\ &+& \pi \omega_{0} \left[{(\Phi - \Phi
  _x - \delta \Phi _x (t))^2\over 2} + {Q^2 \over 2}\right],
\nonumber 
\end{eqnarray}
where $\delta \tPhix (t)$ and $\delta \Phi _x (t)$ are the
fluctuations. If we consider the oscillatory regime, i.e.,
$\tPhix = \pm 1$, we can expand the Hamiltonian around any integer
value of $\Phi = k$, 
\begin{eqnarray*}
  \nonumber
  H &\approx& \mp E_J^0 {\pi \delta \tPhix (t) \over 2} (-1)^k
  \left(1 - {\pi ^2 \over 2} \Phi^2 \right) + \\
  && + \pi \omega _{0} \left[{(\Phi - \Phi _x)^2 \over 2} + {Q^2
  \over 2} \right. \\ && \left. + (\Phi - \Phi _x) \delta \Phi _x (t)
  + {(\delta \Phi 
  _x(t))^2\over 2}\right].
\nonumber 
\end{eqnarray*}
This Hamiltonian contains terms that
describe coupling of the flux $\Phi $ to external flux
fluctuations. Coupling of linear terms in $\Phi $ corresponds to the
amplitude damping of the oscillator, while coupling of quadratic terms
in $\Phi $ is
associated with the phase damping. (Errors
generated by quadratic terms in the variables can be also induced
during the single-qubit operations, as discussed in
Sec.~\ref{sec:logical} - they would also dephase the system.)

Let us start with the amplitude damping of the $LC$ oscillator described by
the following master equation:
\[
  \dot{\rho } = \Gamma \left(a \rho a ^{\dagger} - {1\over 2} a
    ^{\dagger}a \rho - {1\over 2} \rho a ^{\dagger} a \right).
\]  
Here,
\begin{eqnarray*}
  \nonumber
  a &=& {1\over \sqrt{2}}(\Phi + i Q), \\
  a ^{\dagger} &=& {1\over \sqrt{2}}(\Phi - i Q).
\nonumber 
\end{eqnarray*}
For short time intervals $dt$, we may write
\begin{eqnarray}
  \nonumber
  \rho (t + dt) &=& \left(\sqrt{\Gamma dt} a\right)\rho (t)
  \left(\sqrt{\Gamma dt} a ^{\dagger}\right) \\  &+& \left(I - {\Gamma
  \over 2} a 
    ^{\dagger}a dt \right) \rho (t)\left(I - {\Gamma \over 2} a
    ^{\dagger}a dt \right).
  \label{eq:Mege}
\end{eqnarray} 
If we now compare Eqs.~(\ref{eq:Mvie}) and (\ref{eq:Mege}), we find the
Kraus operators to be
\[
  M_1 = \sqrt{\Gamma dt} a, \quad \;   M_2 = I - {\Gamma \over 2} a^{\dagger}a dt.
\]  
The amplitude corresponding to the first operator is \cite{GotPreKit}
\begin{eqnarray*}
  \nonumber
  &&C_1 (\alpha , \beta , dt) = \\
  \nonumber 
  &-& {i \over  2} \left[\delta (\alpha )
    \delta (\beta - \sqrt{\Gamma dt /2}) - \delta (\alpha ) \delta (\beta
    + \sqrt{\Gamma dt/2})\right] \\ 
  &+& {1\over 2} \left[\delta (\alpha -
    \sqrt{\Gamma dt /2}) \delta (\beta ) - \delta (\alpha + \sqrt{\Gamma
      dt / 2}) \delta (\beta ) \right].
\nonumber 
\end{eqnarray*}

The second amplitude $C_2$ is found from
\[
  \int\limits_{}^{} d \alpha d \beta C_2(\alpha ,\beta ,dt) e^{i \beta
    \Phi }e^{ - i \alpha Q} = I - {\Gamma dt\over 2} a ^{\dagger}a,
\]
by applying inverse Fourier transform to both sides of this equation,
which leads to
\begin{eqnarray}
  C_2(\alpha , \beta , dt) &=& \left(1 + {\Gamma dt \over 4}\right)
  \delta (\alpha ) \delta (\beta ) \nonumber\\ \nonumber  &-& {\Gamma dt \over 4}
  \left[{\partial^2\over{\partial \alpha^2 }}\delta (\alpha ) \delta
  (\beta ) + \delta (\alpha ) {\partial^2\over{\partial \beta^2 }} \delta
  (\beta )\right].
\end{eqnarray}
Clearly, after short time intervals, the state can be only slightly
shifted from the code subspace.

The amplitude damping of harmonic oscillator at finite temperatures
leads to the thermal-state solution, i.e., $\rho (\infty ) = e^{-\beta
  H}$. The thermal-state density matrix has all off-diagonal elements
equal to $0$. So, the amplitude damping gives rise indirectly also to
dephasing. However, as already noted, in Josephson-junction systems it
is worth analyzing the pure dephasing mechanism independently.\\
The latter process for the oscillator is described by the master equation
\[
  \dot{\rho } = \Gamma \left[a ^{\dagger} a \rho a ^{\dagger} a -
    {1\over 2} (a ^{\dagger}a)^2 \rho - {1\over 2}\rho (a ^{\dagger}
    a)^2 \right].
\]  
The same procedure like for amplitude damping gives 
\begin{eqnarray*}
  \nonumber
  M_1 &=& \sqrt{\Gamma dt} a ^{\dagger}a \approx {1\over 2i} \left(e^{i
      \sqrt{\Gamma dt} a ^{\dagger}a}-e^{-i \sqrt{\Gamma dt} a
      ^{\dagger}a}\right),\\
  M_2 &=& I - {\Gamma \over 2} (a^{\dagger}a)^2 dt.
\nonumber 
\end{eqnarray*}  
The first operator, $M_1$, is a sum of under-rotation and
over-rotation of the oscillator and has been discussed in
Ref.~\onlinecite{GotPreKit}. $M_2$ is characterized by the  
amplitude
\begin{eqnarray*}
  \nonumber
  &&C_2(\alpha , \beta , dt) = \left(1 - {\Gamma dt\over 8}\right)
  \delta (\alpha )\delta (\beta ) \\
  \nonumber
  &-& {\Gamma dt\over 8} \left\{{\partial^4\over{\partial \alpha^4
        }}\delta (\alpha ) \delta 
    (\beta ) + \delta (\alpha ) {\partial^4\over{\partial \beta^4 }} \delta
    (\beta ) + {\partial^2\over{\partial \alpha^2
        }}\delta (\alpha ){\partial^2\over{\partial \beta^2 }} \delta
    (\beta ) \right. \\ 
\nonumber 
  &&\left.
   - 2 \left[{\partial^2\over{\partial \alpha^2 }}\delta (\alpha ) \delta
  (\beta ) + \delta (\alpha ) {\partial^2\over{\partial \beta^2 }} \delta
  (\beta )\right]\right\},
\end{eqnarray*}
which for small $dt$ should be still sufficiently
localized around the point $(0,0)$ to enable successful error correction.

Regardless of the analysis made, we could also argue that the amplitude $C$ in
Eq.~(\ref{eq:Mtke}) is, in some sense, a wave function. Initially, when
the state is in the code subspace, $C(\alpha , \beta, t = 0) = \delta
(\alpha ) \delta (\beta )$. Any kind of qubit-environment interaction
generates evolution of the function and, if the process is physical,
the wave function will smoothly spread in time. Thus, whatever the sources of
errors, if the state's destruction is not too fast, we should be able
reverse the effects of decoherence. 

In the considered system, we may also reverse some
effects of imperfections in manipulating the system. For instance,
imprecise time measurement is equivalent to over-rotation or
under-rotation, and imprecisely adjusted external flux $\Phi _x$
causes diffusion of the state in the $\Phi$-$Q$ plane.

Error recovery for this class of codes requires two steps
corresponding to two independent errors: shifts in $\Phi $ and $Q$. The
error syndrome (length of the shift) is measured on an ancillary
qubit. First, we prepare the ancilla in the state 
\[
  |\overline{0} \rangle + |\overline{1} \rangle = \sum\limits_{s = - \infty
   }^{\infty} |\Phi = s \rangle .
\]
If the state of the first qubit is shifted by $\alpha $, we have the
initial state of the data qubit and the ancilla in the form
\[
  |\overline{x + \alpha} \rangle (|\overline{0} \rangle + |\overline{1} 
\rangle).
\]
Then, the SUM operation is performed. Since the gate acts on the states 
$|\overline{j} \rangle= \sum_{s} |\Phi = 2s + j \rangle $
like
\[
  \mbox{SUM:} |\overline{j} \rangle |\overline{k} \rangle \rightarrow 
|\overline{j}
  \rangle |\overline{j \oplus k} \rangle,
\]
after the SUM operation, we arrive at
\begin{eqnarray*}
  |\overline{x + \alpha} \rangle (|\overline{x + \alpha } \rangle &+&
    |\overline{x + 1 +
    \alpha} \rangle) \\ &=& |\overline{x + \alpha} \rangle \left(
  \sum\limits_{s = - 
    \infty }^{\infty }| \Phi = s + x + \alpha \rangle\right) .
\end{eqnarray*}
The state of the ancilla is invariant under translation by $1$, and we
can omit $x$ as it can be only $0$ or $1$. The state of the
ancilla contains the information about the shift {\em only} and not about
the actual state of the qubit. By measuring (destructively) the state of
the ancilla system, we read out $\alpha $ mod $1$. If we then shift the
state of the qubit by $\alpha $, the error is corrected. 
The same procedure on the Fourier-transformed state (physically the
operation needs to be delayed by $\pi / 2 \omega _{0}$ - after the
time roles of $\Phi $ and $Q$ are interchanged) yields the
shift in $Q$ that is to be corrected.

\subsection{Requirements to the circuit parameters}

The ingredients required to perform quantum computation and
quantum-error-correction routines have been described here for a
system which 
may be manipulated with great accuracy. However, in real physical systems, it
is rarely the case. Before we conclude our discussion, let us itemize
the main potential problems in manipulating the system. 

First of all, we should briefly comment on the codewords. The states
[Eq.~(\ref{eq:codewords})] are clearly not physical as they contain
components with infinite energy and are non-normalizable. We need to
replace them with approximate, physical codewords. In Ref.~\onlinecite{GotPreKit},
the physical codewords are defined as superpositions of
narrow Gaussians weighted by a Gaussian envelope. If the width of the
narrow peaks [which substitute the delta functions in
Eq.~(\ref{eq:codewords})] is $\kappa $ and we want the states to be
equally localized in $\Phi $ and $Q$, the width of the envelope should
be $\kappa ^{-1}$. In other words, the basis of the (realistic) code
subspace can be defined as
\[
  \nonumber
  |\tilde{j} \rangle \propto \sum_{s = - \infty }^{\infty} e^{-
    \pi ^2 \kappa ^2 (2 s + j)^2 / 2} \int\limits_{}^{} d \Phi  e^{-
    {1\over 2}(\Phi - 2s - j)^2/\kappa 
    ^2} |\Phi  \rangle , 
\]
where $j \in \{0,1\}$. 
There is a finite
overlap of the approximate codewords $|\tilde{0} \rangle$ and
$|\tilde{1} \rangle $,  and the physical states can be
confused during the logical-state readout. However, for a moderate
expectation value 
of the number operator $\langle a ^{\dagger} a \rangle  \approx 16$
($a$ is the annihilation operator of the oscillator),
the error probability is as low as $10^{-6}$ (the estimation of the
error probability due to this overlap has been 
discussed in Ref.~\onlinecite{GotPreKit}). If we require that
the mean energy of the states is an order of magnitude smaller than the
energy gap $\Delta $ (to eliminate dephasing caused by elementary
excitations), we obtain the upper limit of the oscillator frequency
roughly $1$ GHz.

The most challenging in the realization part of the presented procedure may be
encoding. The measurement is assumed to be close to perfect: its
duration should be much shorter than the characteristic
time scale of the oscillator $\tau _m \ll 1/ \omega _{0}$. With
present-day technology, magnetometers with a sensitivity of $10^{-5}\Phi _0
/\mbox{Hz}^{1/2}$ are available. If we want to obtain approximate
codewords in the encoding procedure, we should limit the
oscillator frequency even further, down to a few MHz, which makes the
qubit quite vulnerable to external noise. (The dimensions of the system
need to be increased. Moreover, as the oscillation period gets longer
the errors accumulated during each oscillation become more severe.) 

The next difficulty that appears here is the absolute time
measurement. Once we have prepared the encoded state of the qubit, we need
to keep track of the evolution with accuracy determined by $1/ \omega
_{0}$. This problem can be overcome for a reasonably small number of
operations: even if the oscillator is slightly advanced (or delayed),
we will reset the phase of the oscillator using error-correcting
schemes. However, 
if the number of operations between the error-correcting subroutines
is big, even small imperfections  
may shift the phase so that effective error recovery will be
hardly achievable. 

Finally, the assumed tiny peak separation in the encoded states (of the
order of $\Phi _0$), resulting in the constraint [Eq.~(\ref{eq:condit})]
might require very hardly achievable values of the capacitance and
inductance of the $LC$ oscillator. If we multiply the separation in flux
by any odd integer, we arrive at much weaker conditions. The logical
states $|\bar{0} \rangle $ and $|\bar{1} \rangle $ are still peaked at
maxima and minima, respectively, and the transformation
[Eq.~(\ref{eq:siz})] is still feasible. However, increased separation in
$\Phi $ results in decreased separation of the peaks in the conjugate
variable for which the second single-qubit gate [Eq.~(\ref{eq:six})] does
not work as intended. To overcome this difficulty, it would be necessary to combine
the single-qubit gates with squeezing.

Certainly, if quantum computation is to be of practical meaning, after
the calculations are over we need to read out the state of the
register. Since the logical state is encoded in the flux states, using
the same dc-SQUID as for the purpose of encoding, we may determine
whether the state is closer to $0$ or $1$ (in other words, if the flux is
closer to even or odd value). The potential problems in this procedure
are of the same kind like those discussed above: we need
to perform the measurement when the phase of the oscillator $\omega
_{0} t$ is very close to $0$ (mod $\pi $) or $\pi /2$ (mod
$\pi $) for the
Fourier-transformed state. Also in this case, the time of the
measurement should be much shorter than $1/ \omega _{0}$.

\section{Concluding comments}
\label{sec:comment}

We described a qubit encoded in an infinite-dimensional
system, the rf-SQUID. Utilizing the schemes for quantum error
correction described in Ref.~\onlinecite{GotPreKit}, we showed how to use
the entire Hilbert 
space instead of using simply two states from the spectrum (as it has
been proposed previously for the flux qubits), in this way enabling
error-correcting 
routines with a single rf-SQUID. In principle, one can prepare the logical
state and manipulate the system in a coherent manner so that
universal computation and error correction are possible. The set of gates
consists of symplectic operations, for which amplification of errors
can be avoided,\cite{GotPreKit} and non-symplectic
that can be realized in the system in much more natural way than in
quantum-optical setting. We also
discussed the physical errors and possible difficulties in
experimental realizations. From the latter, we see which procedures
should be further optimized: the encoding accuracy, if independent of the
time necessary to perform projective measurement, would not lead to
very strong constrains on the characteristic time scale of the
system. 

Making use of the procedures presented here, we may consider
also different schemes for quantum error correction in superconducting
nanocircuits, which protect a state of a
wave-packet in an entangled state of many oscillators
(Ref.~\onlinecite{braunstein} and references therein).
Apart from the problem of quantum fault-tolerant computation we may use the
discussed operations to various, not strictly computational schemes,
like unconditional quantum
teleportation of the variables $\Phi $, $Q$, i.e., not only of a
two-dimensional subspace of the system but of its arbitrary wave
function.\cite{braunstein2}

\section*{Acknowledgements}

We are grateful to A.~Shelankov and A.~Shnirman for inspiring
discussions.  This 
work is part of the focused research program ``Quantum information
processing'' of 
the DFG.  Y.M.  was supported by the S.  Kovalevskaya award of the Humboldt
foundation, the BMBF, and the ZIP programme of the German government.

\end{document}